\documentclass[aps,prc,twocolumn,showpacs,superscriptaddress,floatfix,10pt]{revtex4-1}
\usepackage{graphicx}
\usepackage{dcolumn}
\usepackage{bm}
\usepackage{amsfonts}
\usepackage{amssymb}
\usepackage{amsmath}
\begin{document}

\title{Microscopic study of $^{16}$O+$^{16}$O fusion}

\author{C. Simenel}
\affiliation{Department of Nuclear Physics, RSPE, Australian National University, Canberra, ACT 0200, Australia}
\author{R. Keser}
\affiliation{RTE University, Science and Arts Faculty, Department of Physics, 53100, Rize, Turkey}
\author{A.S. Umar}
\author{V.E. Oberacker}
\affiliation{Department of Physics and Astronomy, Vanderbilt University, Nashville, Tennessee 37235, USA}

\date{\today}

\begin{abstract}
We perform a study of $^{16}$O+$^{16}$O fusion at above and below the interaction barrier
energies using three-dimensional time-dependent Hartree-Fock (TDHF)
calculations at above barrier energies and density-constrained TDHF
calculations for the entire energy range. We discuss the variations of the experimental
data at above the barrier energies. Calculations reasonably reproduce the observed energy-dependent
broad oscillations in the fusion excitation functions. These oscillations result from overcoming
$L$-dependent fusion barriers.
The role of the coupling to low-lying octupole states is also discussed.
\end{abstract}
\maketitle

\section{Introduction}

Collisions of light heavy-ions constitute one of the most interesting areas
of low-energy nuclear physics. Scattering of light systems seems to reveal the internal
structure of the colliding nuclei as well as the structure of the composite system
in a profound way. One of the manifestations of this interplay between structure and
reactions is the observation of regular energy-dependent structures in most
scattering cross-sections for collision energies above the Coulomb barrier~\cite{EB85,BT79}.
In conjunction with reaction theory, clustering models~\cite{WO06,IO06}, mean-field studies of
shape isomeric resonances and molecular formations~\cite{BW97,US85,US86,sim13}, and adiabatic time-dependent
Hartree-Fock 
calculations~\cite{GRR83} of fusion barriers have been some of the microscopic approaches
used to study phenomena associated with light systems.

Structures in fusion cross-sections are possible experimental signatures of nuclear molecules~\cite{sch83,leb12}.
However, structures in fusion excitation functions may also appear in light systems which are not necessarily due
to the formation of nuclear molecules.
Such structures or oscillations appear clearly in cross-sections for the fusion of $^{12}$C+$^{12}$C~\cite{spe76b},
$^{12}$C+$^{16}$O~\cite{spe76a}, $^{16}$O+$^{16}$O~\cite{Fernandez78,Tserruya78,Kolata79},
and $^{20}$Ne+$^{20}$Ne~\cite{pof83}.
In particular, the discrete nature of angular momentum may reveal itself in fusion excitation functions as
peaks associated with barriers for specific angular momenta~\cite{van79,pof83,Esb08,Esb12}.

In addition to the intriguing aspects mentioned above, sub-barrier fusion cross-sections
of light systems also carry a significance for astrophysical applications~\cite{Ga04,Ga07,UO12}.
This need for fusion cross-sections at extreme sub Coulomb barrier energies have led to the
discussion of fusion hindrance for such systems~\cite{JR07,das07}.
Recently, Esbensen has provided a comprehensive coupled-channels study of the
$^{16}$O+$^{16}$O fusion for the full energy regime~\cite{Esb08}.
Since most of the data~\cite{Fernandez78,Tserruya78,Kolata79,Wu84,Thomas85,Thomas86,KKT87} for
the $^{16}$O+$^{16}$O system are more than a quarter of a century old, theoretical studies
have been the main tool of recent investigations. As we will discuss below, large variations in the
available fusion data at above-barrier energies have made such studies more difficult.

It is generally acknowledged that the TDHF method provides a useful foundation for a fully
microscopic many-body theory of low-energy heavy-ion reactions~\cite{Ne82,Si12}.
This assumption is predicated in part on the results of fusion excitation calculations
for light-mass systems and particular energy-angle correlation-function calculations
for strongly damped heavy-mass collisions.
Fusion of $^{16}$O+$^{16}$O at above barrier energies was one of the primary testing systems
for early TDHF calculations~\cite{Ko76,CM76,KD77,FKW,BGK,DFFW}, primarily because $^{16}$O
is doubly magic and light systems were easier for computational reasons.
The initial results showed
reasonable agreement with higher energy (around $34$~MeV) fusion data. However, this was
mostly the result of the so-called ``fusion window anomaly'', a non-zero lower angular momentum limit for fusion
due to an unusual degree of transparency for central collisions.
This central deep-inelastic region was not seen experimentally. Later it was shown that
this was primarily due to an approximation made in the effective interaction, the absence of the
spin-orbit term~\cite{USR,RU88}. In these older TDHF calculations, axial symmetry was mostly assumed 
and  lower order discretization techniques were used for numerical implementation.
Non-central impact parameters were often treated via the ``rotating frame approximation''~\cite{Ko76,KD77}.
Today, most TDHF codes employ higher order interpolation methods and operate on a fully three-dimensional
lattice with no unphysical symmetry assumptions~\cite{UO06}.
Naturally, three-dimensional calculations show a higher degree of dissipation in comparison to the
two-dimensional counterparts due to an increased number of degrees-of-freedom sharing the available energy.
In fact, modern TDHF calculations of fusion reactions have reached a good level of description both for light systems
\cite{sim01,UO06,mar06,was08} and heavy systems where fusion hindrance is expected~\cite{Si12,guo12}.
Predictions of reaction dynamics in actinide collisions have also been made recently~\cite{gol09,ked10}.

Almost all TDHF calculations have been done using the Skyrme energy density functional (EDF)~\cite{sky56}.
In addition to the
omission of the spin-orbit term, earlier TDHF calculations also replaced some of the numerically
difficult terms in the Skyrme interaction with a finite-range Yukawa form~\cite{HN77}, without a new fit to the
nuclear properties. Modern Skyrme forces have much improved fit properties to nuclear data that
significantly reduce the variations among them for reproducing global nuclear properties~\cite{rei95,cha98}.
Finally, it is well known~\cite{EB75} that the Skyrme energy density functional also contains time-odd terms
which depend on the spin density, spin kinetic energy density, and the full spin-current pseudotensor.
The time-odd terms vanish for static calculations of even-even nuclei but they should be present
for time-dependent calculations to maintain the Galilean invariance of the collision process~\cite{DD95}.
The Skyrme energy density functional does remain time-reversal invariant
as all the time-odd terms enter in quadratic form or as linear byproducts.
Many of these terms have not been included in older TDHF calculations because of numerical difficulty.
The latest generation of TDHF codes~\cite{kim97,UO06}
used in this study contain all of these time-odd terms.

Fusion at sub-barrier energies is a very challenging problem for nuclear theory.
In particular, it may be extremely sensitive to the internal structure of the collision partners, such as their low-lying collective modes \cite{das98}.
Currently, there is no implementation based on
a true quantum many-body theory of barrier tunneling.
In all sub-barrier fusion calculations one assumes the existence of an ion-ion potential
$V(R)$ which depends on the internuclear distance $R$.
While phenomenological heavy-ion potentials (e.g., Woods-Saxon or double-folding) provide a
useful starting point for the analysis
of fusion data, it is desirable to use a quantum many-body approach which
properly describes the underlying nuclear shell structure of the reaction system.
During the past several years, we have developed the density constrained
time-dependent Hartree-Fock (DC-TDHF) method for calculating
heavy-ion potentials $V(R)$~\cite{UO06a,UO06c,UO09b,UO10a,KU12} which incorporate
all of the dynamical entrance channel effects such as neck formation,
particle exchange, internal excitations, and deformation effects~\cite{UO08a}.

In this paper we carry out TDHF calculations of fusion cross-sections for the $^{16}$O+$^{16}$O
system without making any of the approximations used in earlier calculations. In the next section
we discuss the experimental fusion data. This is followed by above barrier fusion calculations
directly using TDHF. 
Transfer channels and inelastic excitation of octupole states are then discussed. 
Finally, both sub-barrier and above the barrier fusion cross-sections are
calculated using the DC-TDHF approach of obtaining potential barriers from the TDHF time-evolution.

\section{Experimental Data}
\begin{figure}[!htb]
\includegraphics*[width=8.6cm]{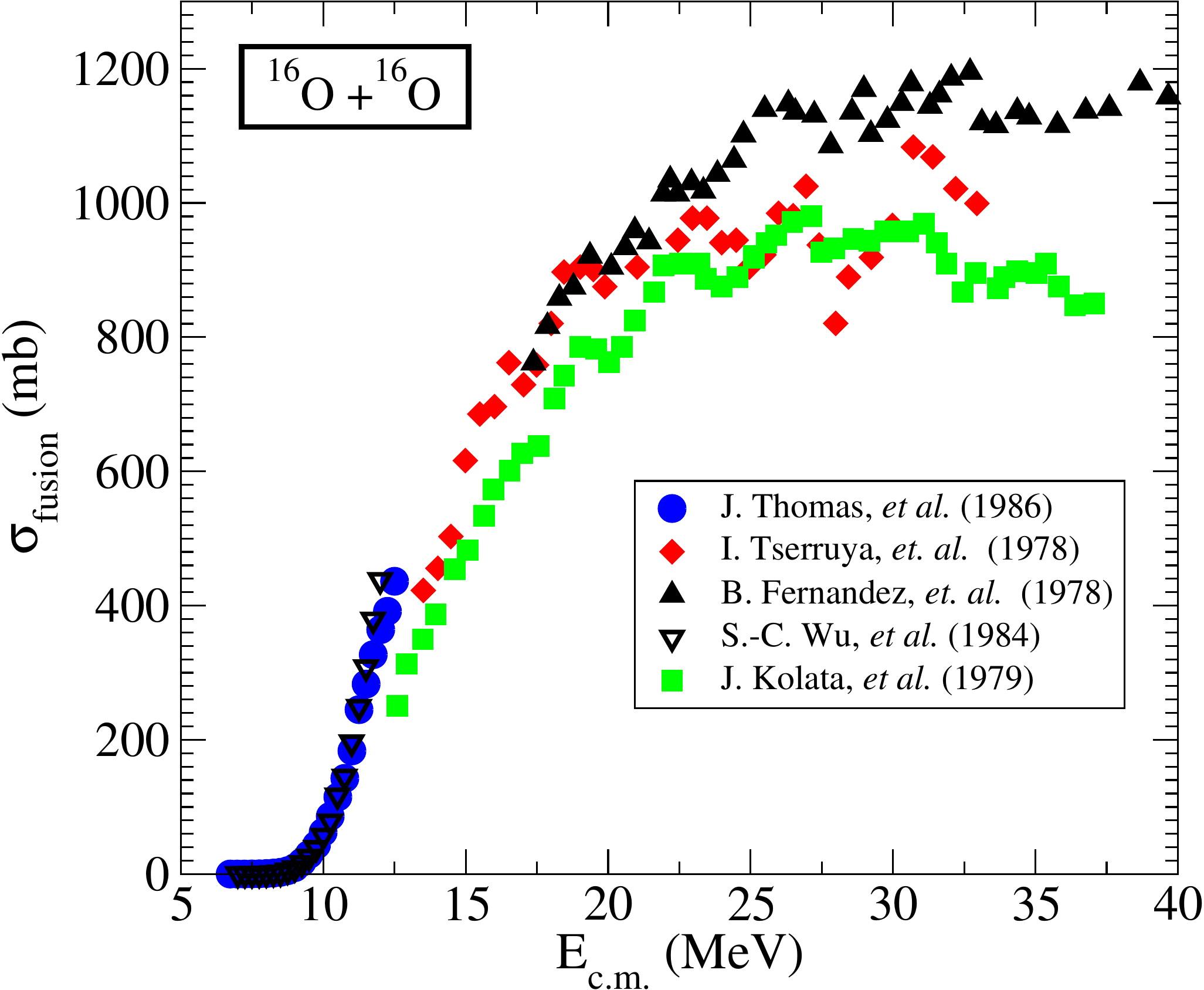}
\caption{(Color online) Experimental data for the $^{16}$O+$^{16}$O fusion excitation functions plotted
on a linear scale.}
\label{fig1}
\end{figure}

During the late 1970s and early 1980s the observed structures in fusion
excitation functions for light-$A$ systems prompted a flurry of experiments measuring
fusion cross-sections for these systems~\cite{EB85}.
Figure~\ref{fig1} shows a selected set of experimental fusion cross-sections for the
$^{16}$O+$^{16}$O system.
The cross-section at higher energies is characterized by broad energy-dependent
oscillations.

Unlike heavy mass systems, for the collisions of light nuclei at bombarding energies
well above the Coulomb barrier the thermalization of the available energy is
much slower thus resulting in collective excitations of the composite system
with a large number of break-up channels.
Consequently, the de-excitation mechanism of the compound system is primarily through the emission of
light-$A$ products as opposed to emission of primary $\gamma$-rays.
The fission mode is assumed to be negligible for light systems.
Broadly, the experimental approach to measuring above barrier fusion
cross-sections for light mass systems falls into three categories.
The first method involves the detection of particle yields originating from fusion products
and is the method used in the earliest experiments~\cite{SP76},
the second method relies on the detection of the de-excitation $\gamma$-rays 
from nuclides produced as evaporation residues (with Kolata \textit{et al.}~\cite{Kolata79} detecting
$12$ nuclides),
and the last method uses
a time-of-flight technique to identify evaporation residues (with Fernandez \textit{et al.}~\cite{Fernandez78}
detecting fragments with $A>20$).
Tserruya \textit{et al.}~\cite{Tserruya78} use both the time-of-flight technique and
a $\Delta E-E$ telescope to identify all products with $Z>10$ or $A>20$ as evaporation residue.
Each one of these methods has its advantages and disadvantages. In addition to their dependence
on the detection system and associated corrections, techniques to identify $\gamma$-rays from
evaporation residue fail to account for transitions to the ground-state and residues formed without
$\gamma$ emission. 
Moreover, high-energy $\gamma$ rays cannot be used to identify nuclei.
The barrier top and sub-barrier fusion data of Wu and Barnes~\cite{Wu84} and Thomas and coworkers~\cite{Thomas85,Thomas86}
are consistent with each other. In these two experiments a lot of attention was paid to degradation
(evaporation) of the target by using gold plated targets and cross-checking the target
thickness with Rutherford Coulomb back-scattering measurements. Among the higher energy
experiments, Fernandez \textit{et al.} use elastic scattering measurements at the beginning
and end of an experimental run to assure that target thicknesses have not changed.
Changing thickness may result in a change in normalization of the cross-sections.
Unfortunately, their data do not extend to lower energies to compare with the lower energy
data. Wu and Barnes undertake a comparison of their results with other experiments
at $E_{\mathrm{c.m.}}=12$~MeV and argue that if Kolata \textit{et al.} data at this
energy are corrected by their summing and branching ratio factor the two results are in
much closer agreement ($438$ versus $481$~mb). Somewhat smaller differences have been
found by Kuronen \textit{et al.}~\cite{KKT87}, who have measured the fusion cross-sections between
$8$ and $14$~MeV and compared their results with the Wu and Barnes and the Thomas data and some of the older data.
Resolution of these differences at all energies is highly desirable.

\section{TDHF Studies of $^{16}$O+$^{16}$O Fusion}

\subsection{Above barrier TDHF calculations}

In this section we present results for $^{16}$O+$^{16}$O fusion by directly using
the \textsc{tdhf3d} code~\cite{kim97} at energies above the Coulomb barrier.
The TDHF equation
\begin{equation}
i\hbar \frac{d\rho}{dt} =\left[h[\rho],\rho\right],
\end{equation}
where $\rho$ is the one-body density matrix of the independent particle system, and $h[\rho]$ is the self-consistent single-particle HF Hamiltonian, is solved iteratively with a time step $\Delta t=1.5\times10^{-24}$~s.
The wave-function is developed on a three-dimensional grid of $56\times28\times28\Delta x^3$, where the mesh grid is $\Delta x=0.8$~fm, and with a plane of symmetry (the collision plane).
More numerical details can be found in Ref.~\cite{Si12}.

The interpretation of
fusion reactions in terms of semi-classical trajectories obtained from the TDHF theory exhibits the
best agreement with experiment for the lightest systems,
since here fusion comprises almost the entire reaction cross section.
Traditionally, the fusion cross-section is given by
\begin{equation}
\sigma_{fus}(E_{\mathrm{c.m.}}) = \frac{\pi\hbar^2}{2\mu E_{\mathrm{c.m.}}} \sum_{L=0}^\infty (2L+1) P_{fus}(L,E_{\mathrm{c.m.}})\;,
\label{eq:cs}
\end{equation}
where $\mu$ is the reduced mass of the system, and $P_{fus}(L,E_{\mathrm{c.m.}})$ is the fusion probability for the partial
wave with orbital angular momentum $L$ at the center-of-mass energy $E_{\mathrm{c.m.}}$.
Due to
the restriction to a single mean-field,
TDHF calculations do not include sub-barrier tunneling of the many-body wave-function, i.e., $P_{fus}^{TDHF}=0$ or~1.
As a result, the fusion cross-section can be estimated with the quantum sharp cut-off formula~\cite{bla54}
\begin{eqnarray}
\sigma_{fus}(E_{\mathrm{c.m.}}) &=& \frac{\pi\hbar^2}{2\mu E_{\mathrm{c.m.}}} \sum_{L=0}^{L_{max}(E_{\mathrm{c.m.}})} (2L+1) \nonumber \\
&=&  \frac{\pi\hbar^2}{2\mu E_{\mathrm{c.m.}}} [L_{max}(E_{\mathrm{c.m.}})+1]^2\;,
\end{eqnarray}
where $L_{max}(E_{\mathrm{c.m.}})$ is the maximum angular momentum at which fusion occurs at $E_{\mathrm{c.m.}}$.
For symmetric systems with $0^+$ ground-states, fusion can only occur for even values of the angular momentum.
The cross-section with the sharp cut-off formula then reads
\begin{figure}[!htb]
\includegraphics*[width=8.6cm]{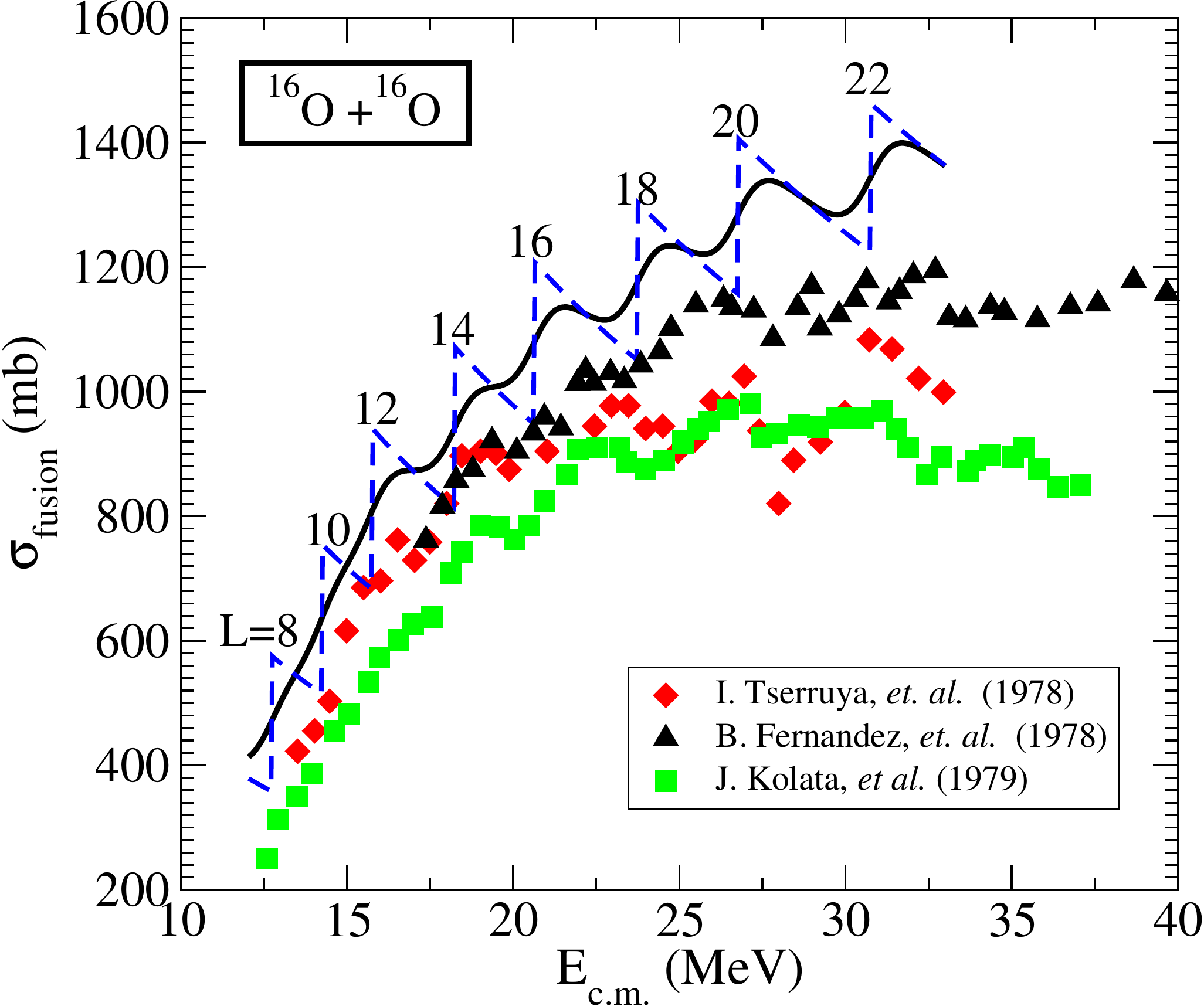}
\caption{\protect (Color online) Above barrier fusion cross-sections as a function of center-of-mass energy in $^{16}$O+$^{16}$O obtained with TDHF
calculations. The cross-sections are computed with the sharp cut-off formula (dashed line) and using Eq.~(\ref{eq:PfusHW})
for the barrier penetration probabilities (solid line). The numbers indicate the position of the barriers $B(L)$.}
\label{fig:csOO}
\end{figure}
\begin{equation}
\sigma_{fus}(E_{\mathrm{c.m.}}) = \frac{\pi\hbar^2}{2\mu E_{\mathrm{c.m.}}} [L_{max}(L_{max}+3)+2]\;.
\label{eq:cseven}
\end{equation}

We have computed fusion cross-sections for the  $^{16}$O+$^{16}$O system as a function of energy 
with the SLy4$d$ Skyrme energy density-functional~\cite{kim97}.
At each center-of-mass energy, a series of runs was performed to pinpoint the ending of fusion or beginning of
the deep-inelastic regime, thus yielding the $L_{max}$ value.
The results of these calculations expressed via
Eq.~(\ref{eq:cseven}) are shown in Fig.~\ref{fig:csOO} with a dashed line.
The sharp increases of the fusion cross-sections at the positions of the angular momentum dependent barriers $B(L)$
are due to the fact that fusion penetration probabilities are either 0 or 1 at the TDHF level.

The sharp edges observed in Fig.~\ref{fig:csOO} can be smoothed by considering tunneling in an approximate way.
As a first approximation, one can estimate the barrier penetration probability according to the Hill-Wheeler
formula~\cite{hil53} with a Fermi function:
\begin{equation}
P_{fus}(L,E_{\mathrm{c.m.}})\simeq \frac{e^{x_L}}{1+e^{x_L}}\;,
\label{eq:PfusHW}
\end{equation}
with $x_L=[E-B(L)]/\varepsilon$.
Choosing the decay constant $\varepsilon=0.4$~\cite{Esb12}, one gets the fusion cross-sections represented by the
solid line in Fig.~\ref{fig:csOO}.

An overall overestimation of the experimental data of Fernandez {\it et al.}~\cite{Fernandez78} by $\sim16\%$ is obtained.
It is interesting to note that the same factor was obtained for the $^{16}$O+$^{208}$Pb system, while the predicted position of the barrier is in excellent agreement with experiment~\cite{Si12}.
Note that this systematic discrepancy remains small, in particular given the fact that the TDHF calculations have no input coming from reaction mechanisms.

Oscillations for $E_{\mathrm{c.m.}}>16$~MeV are clearly visible and due to $L$-dependent barriers with $L\ge12\hbar$.
Note that these oscillations would be less visible for asymmetric systems because all integer values of
$L$ are possible.
In addition, the observation of these oscillations is limited to light systems up to, e.g.,
$^{28}$Si+$^{28}$Si~\cite{gar82,Esb12}.
For heavier systems, the oscillations are indeed expected to be smeared out as the coupling to many reaction
channels sets in~\cite{Esb12}.

\subsection{Role of transfer at sub-barrier energies}

In order to get a deeper insight into the dynamical mechanisms at play in the fusion process, the TDHF approach can also be used to estimate the importance of other channels around the barrier.
Let us first investigate the particle transfer at sub-barrier energies.
Indeed, transfer channels could be a doorway to dissipation and, then, reduce the fusion probability~\cite{eve11}.
The TDHF approach has been used to investigate quasi-elastic transfer reactions in heavy-ion collisions in recent works~\cite{uma08a,was09b,sim10,eve11,yil11,sim12,yab13}.
The proton and neutron transfer probabilities have been computed in $^{16}$O+$^{16}$O at $E_{c.m.}=10$ MeV, i.e., just below the barrier, using the particle number projection technique developed in Ref.~\cite{sim10}.
The resulting probabilities are extremely small, i.e., $p_{1n}\sim p_{1p}\sim 3\times10^{-6}$ which is at the level of the numerical noise.
These results have been confirmed in beyond TDHF calculations where fluctuations at the time-dependent random-phase approximation (TDRPA) level are included (see Refs.~\cite{bal84,sim11,Si12} for details of the technique).
The fact that the transfer probabilities are so small in this reaction is essentially due to the large negative $Q-$values for these channels.
Note that the single-particle wave-functions belonging initially to one nucleus can still be partially transferred to the other fragment~\cite{Si12}.
However, this process, being symmetric, does not change the fragment particle number distributions.
Only transfer to a single-particle state above the Fermi level, which is energetically unfavored,  would induce fluctuations of the particle number probabilities in the fragments.
Note that in these calculations  possible correlations which could enhance, e.g., nucleon pair transfer~\cite{eve11} are neglected.
Beyond TDHF calculations including such correlations should be considered to treat transfer of paired nucleons~\cite{ass09}.
However, pair transfer is unlikely to exceed the transfer of one independent particle.
We then conclude that the sub-barrier fusion in $^{16}$O+$^{16}$O is not affected by transfer channels.

\subsection{Coupling to low-lying octupole states}

\begin{figure}[!htb]
\includegraphics*[width=8.6cm]{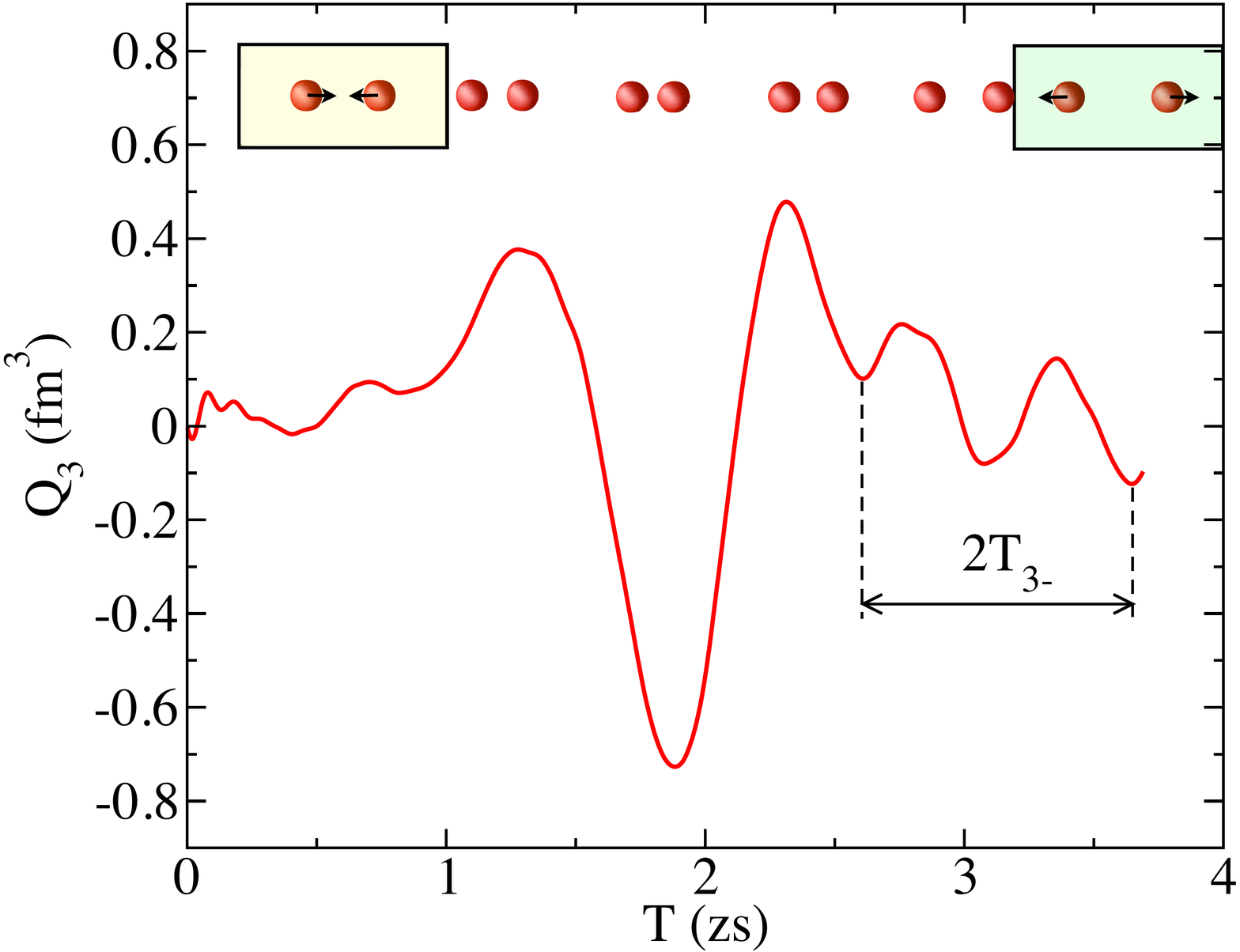}
\caption{\protect (Color online) Time evolution of the octupole moment of the fragment in $x>0$ in an $^{16}$O+$^{16}$O collision at $E_{c.m.}=10$~MeV.
The arrow indicates the time interval $2T_{3^-}$ during which two oscillations of the octupole moment occur.
The snapshots show the isodensity at half the saturation density, $\rho_0/2=0.08$~fm$^{-3}$, at $T=0.6$, 1.2, 1.8, 2.4, 3.0, and 3.6~zs. The numerical box is shown for the first and last snapshots. }
\label{fig:Q3}
\end{figure}

In addition to quasi-elastic transfer, inelastic excitation of low-lying collective states are also known to strongly affect fusion around and below the barrier~\cite{das98}.
Both the coupled-channel approach~\cite{hag12} and TDHF calculations have been used to study the coupling between fusion and rotational motion~\cite{sim04,UO06c} and between fusion and vibrational modes~\cite{sim01,sim07,obe12}.
In the present case, the coupling of the relative motion to the $3^-_1$ octupole phonon in $^{16}$O at $E_{3^-_1}=6.129$~MeV \cite{spe89} may induce a global shift of the barrier to lower energies~\cite{hag97}.
The expectation value of the octupole moment can be computed from the local part of the one-body density using
\begin{equation}
Q_3(t) = \sqrt{\frac{7}{16\pi}}\int d^3r \rho(\mathbf{r}) \left[2\bar{x}^3-3\bar{x}\left(\bar{y}^2+\bar{z}^2\right) \right],
\label{eq:Q3}
\end{equation}
where $\bar{\xi}=\xi-\xi_{c.m.}$.
Because the center of the numerical box in the TDHF calculations is located at $x=y=z=0$,
the octupole moment in one fragment of a central symmetric collision can be obtained by considering the integral in the $x>0$ region in Eq.~(\ref{eq:Q3}) and using the coordinates of the center of mass of the matter distributed in the same region.
The evolution of $Q_3(t)$ of the fragment in the $x>0$ region in an $^{16}$O+$^{16}$O collision at $E_{c.m.}=10$~MeV is shown in Fig.~\ref{fig:Q3}.
Snapshots of the density are also shown at different times.
The increase of $Q_3$ in the approach phase can be interpreted as an effect of Coulomb repulsion inducing an octupole polarization of the fragments.
At short distances between the fragments, the nuclear interaction reverses this polarization and tends to form a neck between the fragments.
After reseparation occurs, the octupole moment clearly oscillates in the exit channel.
The period of this oscillation can be determined from Fig.~\ref{fig:Q3}.
As a result, we get $T_{3^-}\simeq 0.56$~zs.
In the harmonic picture, this oscillation is associated with a vibrational mode at $E_{3^-}=\hbar\frac{2\pi}{T_{3^-}}\simeq7.4$~MeV, which is slightly higher than the experimental value of the $3^-_1$ state.

\begin{figure}[!htb]
\includegraphics*[width=8.6cm]{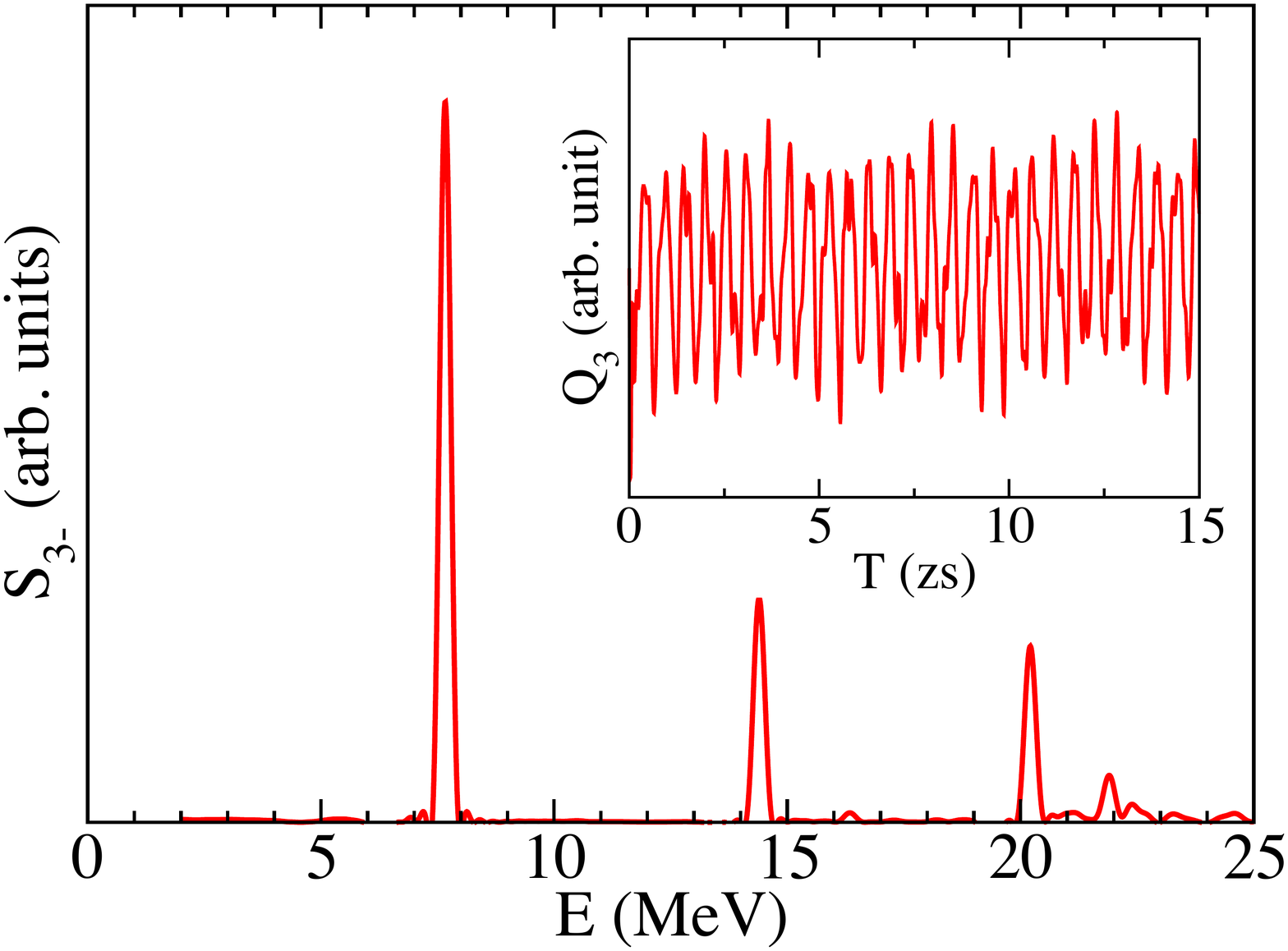}
\caption{\protect (Color online) Strength function of the octupole moment in $^{16}$O obtained from the time evolution of the octupole moment in the linear regime shown in the inset.}
\label{fig:RPA}
\end{figure}

To get a clear assignment of this vibration to the $3^-_1$ phonon, the spectrum of 3$^-$ states has been determined at the RPA level using the same TDHF code with the linear response theory (see Refs.~\cite{ave08,Si12} for similar calculations of bound states and Refs.~\cite{sim03,uma05,mar05,rei07,sim09,fra12,ave13} for unbound states).
The TDHF response to an octupole boost $e^{-i\varepsilon\hat{Q}_3}$ applied on the $^{16}$O Hartree-Fock ground state has been computed.
Note that the initial state is spherical and $Q_3(0)=0$.
The boost velocity $\varepsilon$ is chosen small enough to be in the linear regime, i.e., $Q_3(t)\propto \varepsilon$.
The time evolution of $Q_3(t)$ is shown in the inset of Fig.~\ref{fig:RPA}.
The octupole strength distribution $S_{3^-}(E)$ is determined from $Q_3(t)$ according to
\begin{equation}
S_{3^-}(E)=\frac{-1}{\pi \hbar \varepsilon}\int_0^\infty dt Q_3(t) \sin (Et/\hbar),
\end{equation}
and is shown in Fig.~\ref{fig:RPA}.
A large peak at $E_{3^-}\simeq7.67$~MeV and exhausting $\sim11.2\%$ of the energy weighted sum rule (EWSR) can be seen.
Both the energy and the strength of this peak are of the same order as the $3^-_1$ experimental state (respectively $E_{3^-_1}=6.129$~MeV and $13.1\pm0.6\%$ of the EWSR \cite{spe89}). 
Its energy is close to the energy extracted from the time evolution of the octupole moment of the fragment in Fig.~\ref{fig:Q3}.
The peak at 7.67~MeV in Fig.~\ref{fig:RPA} is   associated with the first phonon of the low-lying octupole vibration in $^{16}$O, and it clearly gives rise to the oscillation observed in the exit channel in Fig.~\ref{fig:Q3}.
Note that other peaks are seen in Fig.~\ref{fig:RPA}.
Several candidates in the spectrum of $3^-$ states in $^{16}$O could be associated with these peaks.
However, their effect on the fusion process is expected to be minor compared to the coupling to the $3^-_1$ phonon, due to their lower strength and larger excitation energy than the $3^-_1$ state~\cite{hag97}.

\begin{figure}[!htb]
\includegraphics*[width=8.6cm]{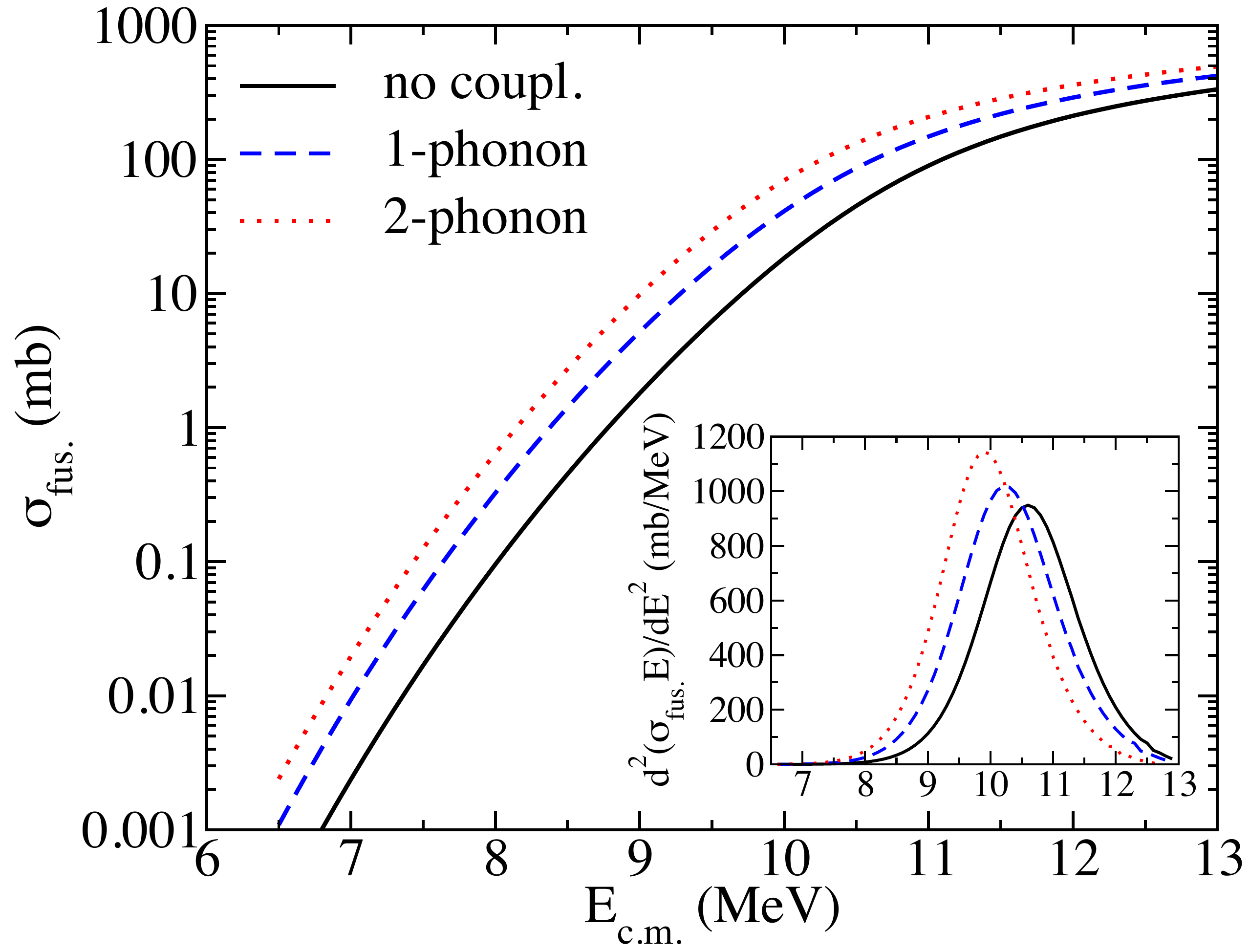}
\caption{\protect (Color online) Fusion cross-sections in the $^{16}$O+$^{16}$O system calculated with the \textsc{ccfull} code with no coupling (solid line), including the coupling to the $3^-_1$ state of one nucleus (dashed line), and including the coupling to the $3^-_1$ state of both nuclei (dotted line). The inset shows the estimate of the barrier distribution from the second derivative of $\sigma_{fus}E$. The parameters of the Woods-Saxon nucleus-nucleus potential and of the coupling are described in the text.}
\label{fig:SP}
\end{figure}

The effect of low-lying collective excitations on fusion cross-sections is usually studied with the coupled-channel approach~\cite{hag12}.
In principle, a microscopic approach such as TDHF could be used to determine the parameters entering coupled channel calculations.
This will be the purpose of a future work.
Here, the effect of the coupling to the $3^-_1$ state on the fusion process is illustrated on Fig.~\ref{fig:SP} with standard coupled channel calculations using the \textsc{ccfull} code~\cite{hag99}.
The nucleus-nucleus potential is a Woods-Saxon potential with a depth $V_0=-65.4$~MeV, a diffuseness $a=0.60$~fm, and a radius parameter $r_0=1.0625$.
These parameters have been fitted to reproduce the Sao-Paulo potential~\cite{cha02} for this system.
The barrier height resulting from this bare potential is $V_B\simeq10.7$~MeV.
The deformation parameter for the nuclear and Coulomb coupling to the ${3^-_1}$ is taken to be $\beta_3=0.733$~\cite{hag97} with the experimental energy $E_{3^-_1}=6.129$~MeV.
We see in Fig.~\ref{fig:SP} that the main effect of the coupling is a global lowering of the barrier by $\sim0.3$MeV per phonon.
When the two phonons are included (one per nucleus), this leads to an enhancement of the sub-barrier fusion by about one order of magnitude.
To conclude, whereas transfer channels can be neglected, we see that the inelastic excitation of low-lying vibrational modes is playing a major role in $^{16}$O+$^{16}$O fusion.

\section{DC-TDHF Studies of $^{16}$O+$^{16}$O Fusion }
The concept of using density as a constraint for calculating collective states
from TDHF time-evolution was first introduced in Ref.~\cite{CR85}, and was used
in calculating collective energy surfaces in connection with nuclear molecular
resonances in Ref.~\cite{US85}. However, its utilization to calculate microscopic
heavy-ion potentials had not been realized until recently~\cite{UO06a}.
In recent years the DC-TDHF method has been applied to calculate fusion barriers
and corresponding cross-sections for over twenty systems.

In this approach we assume that a collective state of the system is characterized only by
the instantaneous TDHF neutron and proton densities.
The lowest static collective energy corresponding to these
densities can be calculated by solving the density-constrained density-functional
problem
\begin{widetext}
\begin{equation}
E_{DC}(t)={\min_{\rho_n,\rho_p}} \left\{ E[\rho_n,\rho_p]+\int d\mathbf{r}\, v_n(\mathbf{r})\left[\rho_n(\mathbf{r})-
\rho_{n}^{tdhf}(\mathbf{r},t)\right]+\int d\mathbf{r}\, v_p(\mathbf{r})\left[\rho_p(\mathbf{r})-
\rho_{p}^{tdhf}(\mathbf{r},t)\right]\right\}\;,
\label{edc}
\end{equation}
\end{widetext}
where $E[\rho_n,\rho_p]$ is the same density-functional used in the TDHF (Skyrme functional) formulation
and additional dependencies have been omitted for notational simplicity, the
quantities $v_{n,p}(\mathbf{r})$ are the Lagrange multipliers, which represent external
fields that constrain the densities during the minimization procedure.
Equation~(\ref{edc}) is equivalent to solving static Hartree-Fock equations subject
to the constraints on neutron and proton densities to remain equal to the instantaneous
TDHF densities while minimizing the energy.

In terms of this state one can write the collective energy as
\begin{equation}
\label{eq:4}
E_{coll}(t)=E_{kin}(\rho(t),\mathbf{j}(t))+E_{DC}(\rho(\mathbf{r},t))\;,
\end{equation}
where the collective kinetic energy $E_{kin}$ is defined as
\begin{equation}
E_{kin} \approx \frac{m}{2}\sum_q\int d^{3}r\; \mathbf{j}^2_q(t)/\rho_q(t)\;,
\end{equation}
with index $q$ being the isospin index for neutrons and protons ($q=n,p$).

This collective energy differs from the conserved TDHF energy only by the amount of
internal excitation present in the TDHF state, namely
\begin{equation}
E^{*}(t)=E_{TDHF} - E_{coll}(t)\;.
\end{equation}
From Eq.~(\ref{eq:4}) it is clear that the density-constrained energy
$E_{DC}$ plays the role of a collective potential. In fact this is
exactly the case except for the fact that it contains the binding
energies of the two colliding nuclei. One can thus define the ion-ion
potential as~\cite{UO06a}
\begin{equation}
V=E_{\mathrm{DC}}(\rho(\mathbf{r},t))-E_{A_{1}}-E_{A_{2}}\;,
\end{equation}
where  $E_{A_{1}}$ and $E_{A_{2}}$ are the binding energies of two nuclei
obtained from a static Hartree-Fock calculation with the same effective
interaction. For describing a collision of two nuclei one can label the
above potential with ion-ion separation distance $R(t)$ obtained during the
TDHF time-evolution. This ion-ion potential $V(R)$ is asymptotically correct
since at large initial separations it exactly reproduces $V_{Coulomb}(R_{max})$,
where $R_{max}=14.5$~fm is the initial distance between the nuclei in the DC-TDHF calculations.
In addition to the ion-ion potential, it is also possible to obtain coordinate
dependent mass parameters. One can compute the ``effective mass'' $M(R)$
using the conservation of energy at zero impact parameter:
\begin{equation}
M(R)=\frac{2[E_{\mathrm{c.m.}}-V(R)]}{\dot{R}^{2}}\;,
\label{eq:mr}
\end{equation}
where the collective velocity $\dot{R}$ is directly obtained from the TDHF evolution and the potential
$V(R)$ from the density constraint calculations. This coordinate dependent mass can be exactly
incorporated into the potential $V(R)$ by using a point-transformation~\cite{UO09b}.

\begin{figure}[!htb]
\includegraphics*[width=8.6cm]{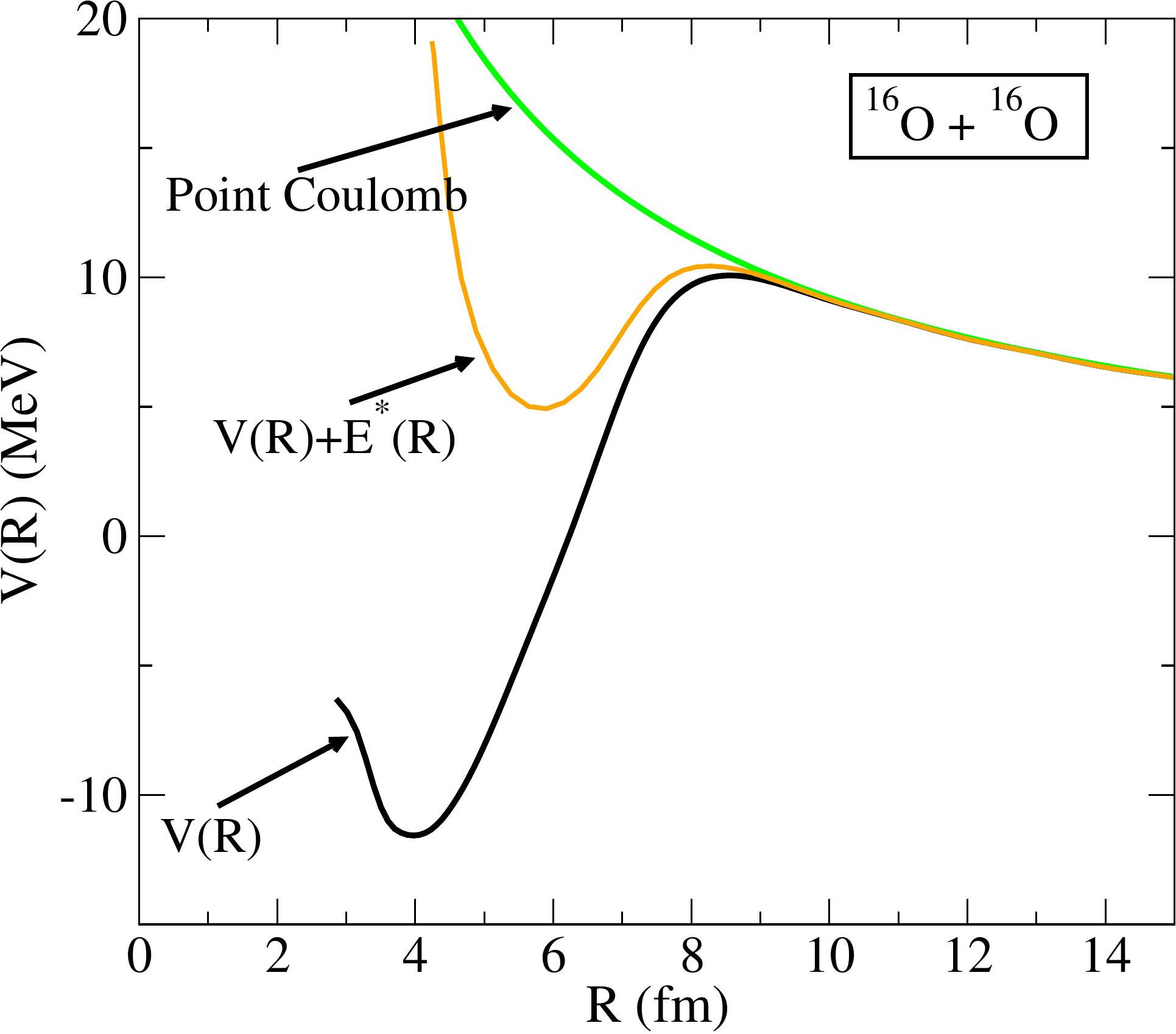}
\caption{\protect (Color online) DC-TDHF ion-ion interaction potential for $^{16}$O+$^{16}$O obtained from TDHF
calculations at $E_{\mathrm{c.m.}}=12$~MeV (black curve). Shown also is a potential which incorporates
the excitation energy at $E_{\mathrm{c.m.}}=20$~MeV (orange curve), as well as the corresponding point-Coulomb
potential (green curve).
}
\label{fig3}
\end{figure}

For the $^{16}$O+$^{16}$O system we have shown~\cite{UO12} excellent agreement between our calculations
and the low-energy sub-barrier data from Refs.~\cite{Thomas85,Thomas86}.
We now extend this work to higher energies to see how our results compare with the available data.
In Fig.~\ref{fig3} we show the calculated DC-TDHF ion-ion potential for the $^{16}$O+$^{16}$O system
evaluated at a TDHF c.m. energy of $12$~MeV. For comparison we also display the corresponding
point-Coulomb potential. The potential barrier has a height of $10.05$~MeV.
For this system there is practically no dependence of the DC-TDHF barrier on the c.m. energy even
if we increase the energy as much as five times the barrier height. This is generally true for light compact
systems whereas for heavy systems a strong energy dependence of DC-TDHF potentials is observed~\cite{UO10a}.
\begin{figure}[!htb]
\includegraphics*[width=8.6cm]{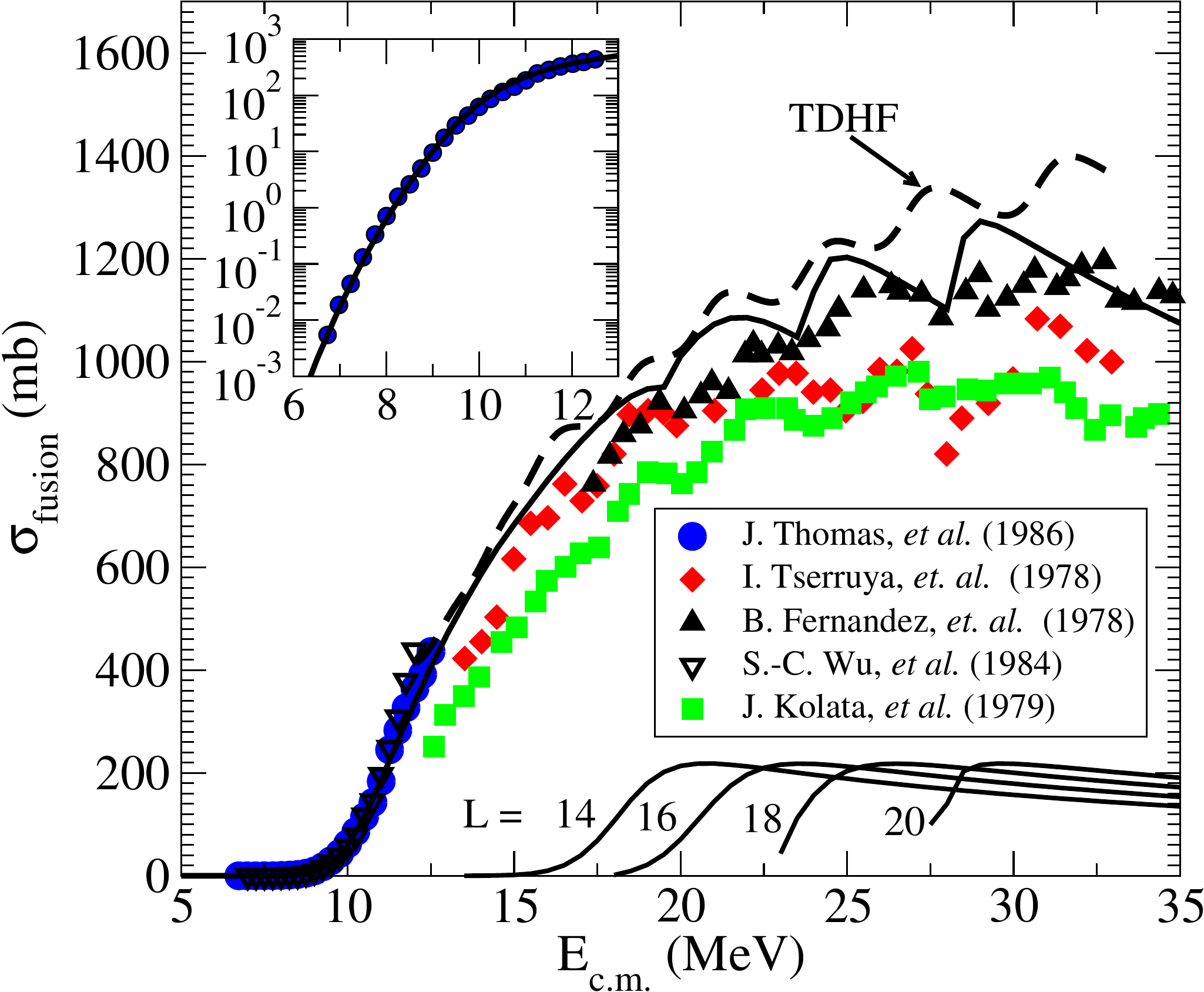}
\caption{\protect (Color online) Fusion cross-sections for $^{16}$O+$^{16}$O obtained from the DC-TDHF potential
shown in Fig.~\ref{fig3} (black curve, calculated at $E_{\mathrm{c.m.}}=12$~MeV) compared with the TDHF calculations (dashed line) and with the experimental data.
The lower curves show the contributions arising from the opening of new orbital angular momentum channels.
\label{fig4}}
\end{figure}

Figure~\ref{fig4} shows the fusion cross-sections corresponding to the DC-TDHF barrier shown in Fig.~\ref{fig3}
(black solid curve) compared with the experimental data on a linear scale. The inset shows
a logarithmic plot of the low-energy fusion cross-sections
which are in excellent agreement with the experimental data~\cite{Thomas85,Thomas86}.
It is interesting to note that the gross oscillations in the cross-section at higher energies
are correctly reproduced in our calculations. This is simply due to opening of new $L$-channels
as we increase the collision energy. Individual contributions to the cross-section
from higher $L$ values are also shown on the lower part of the  plot.
Cross-sections are calculated by directly integrating the
Schr\"odinger equation
\begin{equation}
\left [ \frac{-\hbar^2}{2\mu} \frac{d^2}{d{R}^2} + \frac{\hbar^2 \ell (\ell+1)}{2 \mu {R}^2} + V({R})
 - E_\mathrm{c.m.} \right] \psi_{\ell}({R}) = 0 \;,
\label{eq:Schroed1}
\end{equation}
using the well-established
{\it Incoming Wave Boundary Condition} (IWBC) method~\cite{HW07} to obtain
the barrier penetrabilities $P_{fus}(L,E_{\mathrm{c.m.}})$ which determine the total fusion cross
section
[Eq.~(\ref{eq:cs})].
We observe that while the calculated cross-sections are in excellent agreement with the data
of Fernandez~\textit{et al.}~\cite{Fernandez78}, they are higher by as much as $25$\% than
the lowest data points at $E_{\mathrm{c.m.}}=30$~MeV.
However, one should keep in mind that the DC-TDHF potential was obtained by using a parameter-free
microscopic theory.

The TDHF results of Fig.~\ref{fig:csOO} have also been reported on Fig.~\ref{fig4} (dashed line).  
Although a good agreement with the DC-TDHF results is obtained at energies $E<25$~MeV,
the TDHF fusion cross-sections are larger at higher energies. 
Detailed examination shows that this discrepancy is 
due to the earlier closing of the $L=20$ window
and the total absence of the $L=22$ partial wave in the
potential model calculations of the DC-TDHF formulation. 

In the DC-TDHF approach only the dynamical density evolution for central collision is used in the calculation of the potential. 
Then the cross-sections are determined using the isocentrifugal approximation. 
This is naturally an approximation to the full many-body calculation but seems to be a very good approximation at lower energies and allows the calculation
of sub-barrier cross-sections, where only small angular momenta $L$ contribute.
At the higher energies discussed in this paper (up to three times the barrier height),
the fusion cross-section is determined by larger values of the critical angular momentum.
The smaller potential pocket at high $L$ may lead to a higher sensitivity of the fusion cross-section to the details of the potential. 
The observed discrepancy would indicate a breakdown of the isocentrifugal approximation, meaning that the TDHF potential at high $L$ is not exactly the one at $L=0$ plus a centrifugal potential. 
Naturally, direct TDHF
calculations contain all of the dynamics and therefore should
be more reliable at these energies.

\begin{figure}[!htb]
\includegraphics*[width=8.6cm]{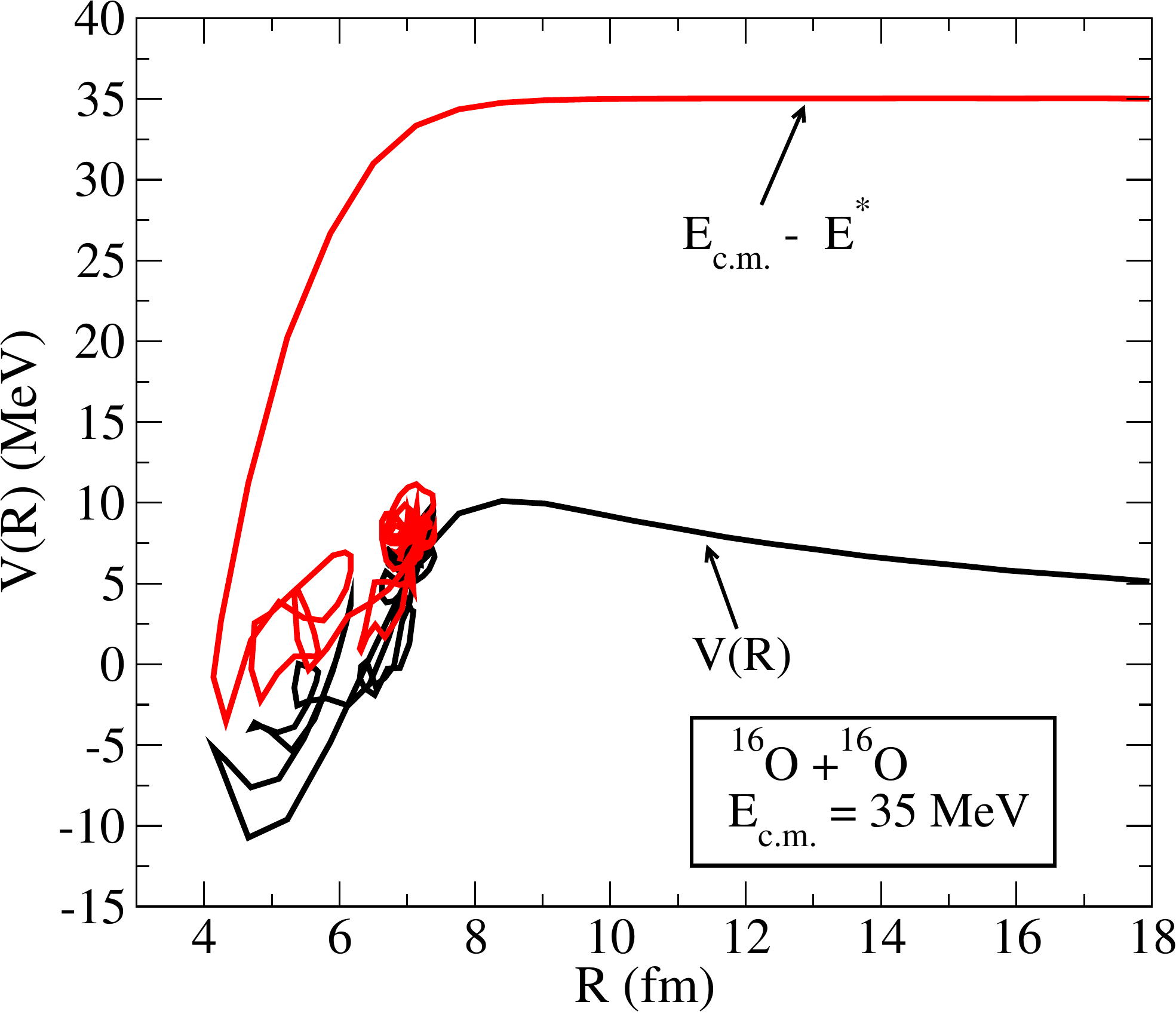}
\caption{\protect (Color online) Long-time evolution of the ion-ion potential, $V(R)$, and the excitation energy,
$E_{\mathrm{c.m.}}-E^{*}$,
for the head-on collision of the $\mathrm{^{16}O}+\mathrm{^{16}O}$ system at a collision energy of
$35$~MeV as a function of the ion-ion distance $R$.}
\label{fig5}
\end{figure}

The reactions of light systems at high energies (two to four times the barrier height) is complicated
both experimentally and theoretically due to the presence of many breakup channels and excitations.
We expect the TDHF results to yield a higher fusion cross-section since many of the
breakup channels are not naturally available in TDHF and they will appear as fusion.
However, a closer investigation of the TDHF dynamics and the microscopically calculated
excitation energy clearly indicates that a significant portion of the collective kinetic
energy is not equilibrated.
In Fig.~\ref{fig5} we show the long-time evolution of the DC-TDHF potential $V(R)$ together
with the quantity $E_{\mathrm{c.m.}}-E^{*}$ calculated at the TDHF energy $E_{\mathrm{c.m.}}=35$~MeV.
We see that after the system traverses through the first potential minimum it starts to rise
and go through successive higher minima and finally settles almost near the top of the entrance
channel barrier. This is perhaps more clearly seen through the evolution of $E_{\mathrm{c.m.}}-E^{*}$,
which is some measure of dissipation. This behavior occurs because the excitation
energy, $E^{*}$, is not distributed in an irreversible fashion but a certain fraction of it
seems to be reversible, going into collective modes. In a fully quantal calculation this may
partially lead to a break-up channel. The situation for the collision at $E_{\mathrm{c.m.}}=12$~MeV
is very different. In this case the system settles close to the first barrier minimum.

\section{Conclusions}

A microscopic study of the $^{16}$O+$^{16}$O fusion reaction has been performed.
Available experimental data sets above the barrier clearly disagree with each other.
Calculations with the TDHF and the DC-TDHF methods are in better agreement with the data exhibiting the largest cross-sections.
The oscillations of the cross-sections are interpreted as an effect of overcoming angular momentum dependent barriers.
The sub-barrier cross-sections are  very well reproduced by the DC-TDHF calculations.

The DC-TDHF method has also been used to investigate the dissipative mechanisms.
The latter rapidly set in inside the barrier, converting the kinetic energy of the relative motion into excitation energy of the fragments.
In central collisions slightly above the barrier the system then settles close to the first barrier minimum.
In contrast, well above the barrier the system settles near the top of the entrance channel barrier into a di-nuclear configuration, possibly leading to a break-up channel.

The effect of coupling to other channels has been studied.
Although the transfer reactions at sub-barrier energies are shown to be negligible, an oscillation of the octupole moment of the fragment in the exit channel indicates a coupling between the relative motion and the $3^-_1$ states in $^{16}$O.
Coupled-channel calculations show that this coupling shifts the barrier to lower energy, increasing the sub-barrier fusion cross-section by about one order of magnitude.

Due to its small number of constituents, and to the magic nature of the collision partners, the $^{16}$O+$^{16}$O system is an ideal benchmark for low energy reaction theory.
Future theoretical models able to treat in a fully microscopic manner the reaction mechanisms from deep sub-barrier energies to well above the barrier should be tested on this system.
However, reliable experimental data above the barrier are highly desirable.

\begin{acknowledgments}

Useful discussions with M. Dasgupta and D. J. Hinde are acknowledged.
M. Dasgupta is also warmly thanked for her help with the coupled channel calculations.
This work has been supported by the U.S. Department of Energy under Grant No.
DE-FG02-96ER40975 with Vanderbilt University,
 and by the Australian Research Council under the Future Fellowship FT120100760, Laureate Fellowship FL110100098 and Discovery Grant DP1094947.
 Part of the calculations have been performed on the NCI National Facility in Canberra, Australia, which is supported by the Australian Commonwealth Government.

\end{acknowledgments}

\end{document}